\documentclass[ reprint,nofootinbib,amsmath,amssymb,onecolumn,aps]{revtex4-2}
\usepackage{appendix}
\usepackage{graphicx}
\usepackage{dcolumn}
\usepackage[colorlinks,linkcolor=magenta,anchorcolor=cyan,citecolor=blue]{hyperref}
\usepackage{physics}
\usepackage{bm}

\def\be{\begin{equation}}
\def\ee{\end{equation}}
\def\ba{\begin{eqnarray}}
\def\ea{\end{eqnarray}}

\def\dbar{{\mathchar '26\mkern -10mu\delta}}

           \def\X {\Xi}   \def\.{\cdot}

\begin{document}
\title{Higher derivative corrections to Kerr-AdS black hole thermodynamics}
\author{Wei Guo$^{1,2}$}
\email{guow@mail.bnu.edu.cn}
\author{Xiyao Guo$^{1,2}$}
\email{xiyaoguo@mail.bnu.edu.cn}
\author{Xin Lan$^{1,2}$}
\email{xinlan@mail.bnu.edu.cn}
\author{Hongbao Zhang$^{1,2}$}
\email{hongbaozhang@bnu.edu.cn}
\author{Wei Zhang$^{1,2}$}
\email{w.zhang@mail.bnu.edu.cn}
\affiliation{$^1$School of Physics and Astronomy, Beijing Normal University, Beijing 100875, China\label{addr1}\\
$^2$ Key Laboratory of Multiscale Spin Physics, Ministry of Education, Beijing Normal University, Beijing 100875, China}

\date{\today}

\begin{abstract}
Instead of the much more involved covariant counterterm method, we apply the well justified background subtraction method to calculate the first order corrections to Kerr-AdS black hole thermodynamics induced by the higher derivative terms up to the cubic of Riemann tensor, where the computation is further simplified by the decomposition trick for the bulk action. The validity of our results is further substantiated by examining the corrections induced by the Gauss-Bonnet term. Moreover, by comparing our results with those obtained via the ADM and Wald formulas in Lorentzian signature, we can extract some generic information about the first order corrected black hole solution induced by each higher derivative term. 
%With this, we further establish that the background subtraction method is applicable not only to Einstein's gravity, but also to its higher derivative corrections for black holes in both asymptotically flat and AdS spacetimes. This thus corrects the longstanding bias that the background subtraction method is rather restrictive compared to the alternative covariant counterterm method.
%As a bonus, we also obtain the corresponding expression of the Gibbs free energy in terms of the Euclidean on-shell action. This can be regarded as the first proof of such a relation, which has been assumed to be also valid beyond Einstein's gravity without derivation before. 

%By working with the covariant phase space formalism, we have shown that not only can the Hamiltonian conjugate to a Killing vector field $\xi$ be expressed as the sum of the associated Noether charge and $\xi$ contracted with the Hilbert action boundary term for $F(R_{abcd})$ gravity, but also be written as its contraction with another $\xi$ independent tensor field. With this, we have proven the equivalence of Noether charge and Hilbert action boundary term formulae for the stationary black hole entropy in $F(R_{abcd})$ gravity, which is further substantiated by our explicit computation using both formulae. 
\end{abstract}
\maketitle
%\onecolumngrid
\section{Introduction}

In the Euclidean approach to quantum gravity, there are two available methods in calculating the Gibbs free energy of stationary black holes and the related thermodynamic quantities. One is called the background subtraction method proposed by Hawking and his companions\cite{Hartle,GH,HP,HH,GPP}, and the other is called the covariant counterterm method inspired by AdS/CFT correspondence\cite{BFS,PS,MM}. Both methods are believed to have their respective disadvantages. The covariant counterterm method generically needs notoriously complicated terms added on the boundary on top of the generalized Gibbons-Hawking-York term for a general diffeomorphism covariant theory of gravity, while the background subtraction method is believed to be rather restrictive in the sense that the induced metric of the stationary black holes on the boundary cannot always be isometrically embedded into the reference spacetime\cite{PS,MM}. However, by applying the covariant phase space formalism in an innovative manner, it has been shown most recently in \cite{HB} that the criterion for the applicability of the background subtraction method does not require the aforementioned isometric embedding at all, instead the background subtraction method turns out to be as applicable as the covariant counterterm method indeed. Speaking specifically, not only is the applicability of the background subtraction method well established for Einstein's gravity, but also for its higher derivative corrections in both asymptotically flat and anti-de Sitter (AdS) spacetimes. Accordingly, the much simpler background subtraction method emerges naturally as the favorable one to be employed in the ongoing effort for the investigation of the higher derivative corrections to black hole thermodynamics, whereby not only can we obtain some universal information about the UV behavior of quantum gravity in the context of effective field theory, but also gain some insight into the dynamics of the boundary quantum systems at a finite coupling and a finite $N$ through the lens of AdS/CFT corespondence.
With this in mind, in this paper we shall apply the background subtraction method to calculate the first order corrections to the four dimensional Kerr-AdS black hole thermodynamics by the higher derivative terms up to the cubic of Riemann tensor, which, to our knowledge, remains unexplored,
%\footnote{The background subtraction method has been used to obtain an explicit expression for the higher derivative corrections to the four dimensional Kerr black hole thermodynamics in \cite{RS}, where it is briefly claimed without an explicit demonstration in the discussion section that the corresponding result for the Kerr-AdS black hole has also been obtained in the same way. However, as shown in \cite{HB} as well as in the present paper, the relevant calculation associated with the Kerr-AdS black hole is not a trivial generalization of that associated with the Kerr black hole.}
although much effort has recently been made in the investigation of the higher derivative corrections to thermodynamic quantities for various black holes\cite{RS,Cheung,Cremonini,Belgium,Melo,Bobev,Xiao1,Cassani,Noumi,Cassani1,Ma,Zatti,Hu,Ma2,Cano,Massai,Cassani2,Xiao2}. To achieve this, we shall also facilitate our calculation by adopting the strategy developed in \cite{Xiao2}, with which we can extract some generic features about the first order corrected black hole solution, although we are not required to solve it explicitly at all. 

%But nevertheless, as evidenced by the final correct result it gives rise to,  the background subtraction method seems also applicable to the circumstances where such an embedding is impossible.

%Among others, one main purpose of this paper is to examine the applicability of the background subtraction method by working with the covariant phase space formalism. As a result, we succeed in identifying the necessary and sufficient condition for the validity of the background subtraction method. Then we further show that the resulting criterion is satisfied not only by Einstein's gravity but also by its higher derivative corrections for both black holes in asymptotically flat and AdS spacetimes, even though the induced metrics of the stationary black holes and the reference spacetime are not exactly the same on the boundary. With this in mind,  our result, on the one hand, offers a priori rather than a posteriori justification for the applicability of the background subtraction method in the aforementioned more generic circumstances. Accordingly, the background subtraction method is supposed to be as applicable as the covariant counterterm method. With this in mind, we further apply the background subtraction method to calculate the higher derivative corrections to Kerr-AdS black hole thermodynamics, where we find that the relevant computation can be simplified greatly by resorting to the spinor decomposition of Weyl tensor. 

The structure of this paper is organized in the following way. In the next section, we shall provide a brief review of the applicability of the background subtraction method to asymptotically AdS spacetimes, where we also work with a different coordinate system from that in \cite{HB} to demonstrate that the criterion for the applicability of the background subtraction method is satisfied by Einstein's gravity as it should be the case. Then in Section \ref{33}, we shall apply the viable background subtraction method to calculate the first order corrected Gibbs free energy as well as the associated entropy, angular momentum and mass in the grand canonical ensemble. By comparing the resulting angular momentum and mass with those obtained via the Arnowitt-Deser-Misner (ADM) formula as well as the resulting entropy with that obtained via the Wald formula, we further extract some generic information about the full first order corrected black hole solution. Moreover, we substantiate the validity of our result by examining the corrections induced by the Gauss-Bonnet term.  We shall conclude our paper with some discussions in the last section. 

But before proceeding, we like to summarize what will be accomplished in this work that
is novel as follows. First, compared to the coordinate system used in \cite{HB}, we adopt a more convenient one to justify the applicability of the background subtraction method to Einstein's gravity in asymptotically AdS spacetimes. Second, with the much simpler background subtraction method, we are the first one to achieve the first order corrections to the four dimensional Kerr-AdS black hole thermodynamics by the higher derivative terms. Last but not least, simply by comparing our results with those obtained from the ADM and Wald formulas, we find that one can also succeed in extracting some generic information about the full first order corrected black hole solution by each higher derivative term without solving it out. 

%We follow the notation and conventions of \cite{GR}. In particular, we shall use the boldface letters to denote differential forms with the tensor indices suppressed. 

%Among others, one main purpose of this paper is to examine the applicability of the background subtraction method by working with the covariant phase space formalism. As a result, we succeed in identifying the necessary and sufficient condition for the validity of the background subtraction method. Then we further show that the resulting criterion is satisfied not only by Einstein's gravity but also by its higher derivative corrections for both black holes in asymptotically flat and AdS spacetimes, even though the induced metrics of the stationary black holes and the reference spacetime are not exactly the same on the boundary. With this in mind,  our result, on the one hand, offers a priori rather than a posteriori justification for the applicability of the background subtraction method in the aforementioned more generic circumstances. Accordingly, the background subtraction method is supposed to be as applicable as the covariant counterterm method. With this in mind, we further apply the background subtraction method to calculate the higher derivative corrections to Kerr-AdS black hole thermodynamics, where we find that the relevant computation can be simplified greatly by resorting to the spinor decomposition of Weyl tensor. 

\section{Applicability of background subtraction method to asymptotically AdS spacetimes }
In this section, we shall first review the criterion for the applicability of the background subtraction method in $F(R_{abcd})$ gravity. For details, please refer to \cite{HB}. Then by taking another different coordinate system from that in \cite{HB}, we shall show that the criterion is satisfied by Einstein's gravity in asymptotically AdS spacetimes, which can be regarded as a double check of the claim made in \cite{HB}. We conclude this section by recapping the main statement in \cite{HB} regarding the applicability of the background subtraction method to the higher derivative gravity as the corrections to Einstein's gravity within the framework of effective field theory. 

Let us start from the following Lagrangian form for $F(R_{abcd})$ gravity, i.e.,
\begin{equation}
\mathbf{L}=\bm{\epsilon}F(R_{abcd},g_{ab})
\end{equation}
 with $\bm{\epsilon}$ the spacetime volume and $F$ an arbitrary function of the Riemann tensor $R_{abcd}$ and the metric $g_{ab}$. Its variation further gives rise to 
\begin{equation}\label{variationradius}
    \delta \mathbf{L}= \bm\epsilon E_g^{ab}\delta g_{ab}+d\mathbf{\Theta}.
\end{equation}
 Here
\begin{eqnarray}
E_g^{ab}=\frac{1}{2}g^{ab}F+\frac{1}{2}\frac{\partial F}{\partial g_{ab}}+2\nabla_{c}\nabla_{d}\psi^{c(ab)d}
\end{eqnarray}
with $E_g^{ab}=0$ the equation of motion,
and $\mathbf{\Theta}=\theta\cdot\bm{\epsilon}$ is the bulk symplectic potential with 
 \begin{equation}
\theta^a=2(\nabla_d\psi^{bdca}\delta g_{bc}-\psi^{bdca}\nabla_d\delta g_{bc}),
\end{equation}
where $\psi^{abcd}$ is defined as the derivative of $F$ with respect to $R_{abcd}$ by assuming it is independent of the metric, namely
$\psi^{abcd}\equiv \frac{\partial F}{\partial R_{abcd}}$.

With the outward-pointing normal vector and the induced metric of a timelike boundary $\Gamma$ denoted respectively as $n_a$ and $h_{ab}$, the variation of the metric can be expressed as follows
\begin{equation}
    \delta g^{ab}|_\Gamma=-2\delta an^an^b-\dbar{A}^an^b-\dbar{A}^bn^a+\delta h^{ab}
\end{equation}
with $n_a\dbar{A}^a=0$, where we have required $\Gamma$ remain fixed under variation such that $\delta n_a=\delta a n_a$.
Accordingly, $\mathbf{\Theta}$ can be rewritten in the following form
\begin{eqnarray}\label{boundaryex}
    \mathbf{\Theta}|_{\Gamma}=-\delta \mathbf{B}+d\mathbf{C}+\mathbf{F}.
\end{eqnarray}
Here
\begin{eqnarray}
    \mathbf{B}=4\Psi_{ab}K^{ab}\hat{\bm\epsilon},\quad
    \mathbf{C}=\mathbf\omega\cdot\hat{\bm\epsilon},\quad
    \mathbf{F}=\hat{\bm\epsilon}(T_{hbc}\delta h^{bc}+T_{\Psi bc}\delta\Psi^{bc})
\end{eqnarray}
with $K_{ab}$ the extrinsic curvature, $\hat{\bm\epsilon}$ the induced volume defined as  $\bm\epsilon=\mathbf n\wedge \hat{\bm\epsilon}$, $\Psi_{ab}=\psi_{acbd}n^cn^d$, and
\begin{eqnarray}\label{bc}
    \omega^a&=&-2\Psi^a{}_b\dbar A^b+2h^{ae}\psi_{ecdb}n^d\delta h^{bc},\nonumber\\
    T_{hbc}&=&-2\Psi_{de}K^{de}h_{bc}+2n^a\nabla^e\psi_{deaf}h^d{}_{(b}h^f{}_{c)}-2 \Psi_{a(b}K^a{}_{c)}-2D^a(h_a{} ^eh_{(c}{}^f\psi_{|efd|b)}n^d),\nonumber\\
    T_{\Psi bc}&=&4 K_{bc},
\end{eqnarray}
where $D_a$ denotes the covariant derivative on $\Gamma$.

Now let $\Sigma$ be a spacelike hypersurface emanating from the birfurcation surface $\mathcal{B}$ of a stationary black hole and terminating at $\Gamma$ with their intersection denoted by $\mathcal{S}$. Then one can show that 
if and only if 
    \begin{equation}\label{ourcriterion}
        \int_{\mathcal{S}}\xi\cdot[\mathbf{F}]=0, \quad  \texttt{as} \quad \mathcal{S}\rightarrow \texttt{spatial infinity}
    \end{equation}
with the square bracket representing the difference from the reference spacetime as $[\mathbf{F}]\equiv \mathbf{F}-\mathbf{F}^0$, the first law of black hole thermodynamics holds, i.e.,
   \begin{equation}
     T\delta S=\delta [H_\xi]
    \end{equation}
    with the Hawking temperature $T=\frac{\kappa}{2\pi}$.
    Here $\xi$ is the Killing vector field normal to the black hole event horizon and tangential to $\Gamma$. The black hole entropy is given by
    \begin{equation}\label{entropyformula}
        S=\beta\int_\mathcal{B}\mathbf{Q}_\xi
        \end{equation}
     with $\beta=\frac{1}{T}$ the period along the Euclidean time $\tau=it$,  and the Noether charge
    \begin{equation}
\mathbf{Q}_\xi=-\psi^{cadb}\nabla_{[d}\xi_{b]}\bm{\epsilon}_{ca\cdot\cdot\cdot}.
\end{equation}
      The Hamiltonian conjugate to $\xi$ can be expressed as
    \begin{equation}
    H_\xi=\int_\mathcal{S}q_\xi\cdot\hat{\bm{\epsilon}},
    \end{equation}
    where 
    \begin{equation}
        q_\xi^a=\mathcal{T}^a{}_c\xi^c
    \end{equation}
    with the generalized Brown-York boundary energy momentum tensor 
    $\mathcal{T}^a{}_c= -(2T_h{}^a{}_c+2\Psi^{ab}T_{\Psi cb})$.
    
    %as $\Gamma$ approaches the spatial infinity,
     The condition (\ref{ourcriterion}) together with the finiteness condition for $[H_\xi]$ gives rise to the criterion for the applicability of the background subtraction method. If this criterion is fulfilled, the Gibbs free energy of the stationary black hole can be expressed as %by going to the Euclidean sector as follows. First, by Eq. (\ref{Noethercurrent}), we have 
    %\begin{equation}\label{Gibbs2}
      %  \beta\int_\infty \mathbf{Q}_\xi-S=-\beta\int_\Sigma\xi\cdot \mathbf{L}=\int_M d\tau\wedge \frac{\partial}{\partial\tau}\cdot\mathbf{L}_E=\int_M\mathbf{L}_E,
    %\end{equation}
   % where $\beta=\frac{2\pi}{\kappa}=\frac{1}{T}$ denotes the period along the Euclidean time $\tau=it$ and $\mathbf{L}_E=-i\mathbf{L}$. Similarly, we also have
    %\begin{equation}\label{Gibbs1}
    %\beta\int_\infty \mathbf{Q}^0_\xi=\int_{M^0}\mathbf{L}^0_E
    %\end{equation}
    %Combining it with Eq. (\ref{Gibbs3}) as well as Eq. (\ref{Gibbs2}) and using the fact $X_\xi\cdot \mathbf{C}=0$, we end up with the expression of the Gibbs free energy as
    \begin{equation}
        \beta G\equiv\beta([H_\xi]-TS)=[I_E].
    \end{equation}
    Here the Euclidean action $I_E$ is given by
    \begin{equation}
        I_E=\int_M\mathbf{L}_E+\int_\infty \mathbf{B}_E
    \end{equation}
    with $\mathbf{L}_E=-i\mathbf{L}$ and $\mathbf{B}_E=-i\mathbf{B}$.
    %, where the sign in front of the boundary integral comes from our previous convention for the orientations on $\Gamma$ and $\partial \Sigma_1$.
    %It is noteworthy that the aforementioned finiteness condition for $[H_\xi]$ is equivalent to the finiteness condition for $[I_E]$, which is satisfied if and only if 
   % \begin{equation}\label{finiteness}
    %    \int_{\Gamma_E}\mathbf{F}_E^0
  %  \end{equation}
   % is finite as the Euclidean counterpart $\Gamma_E$ of $\Gamma$ approaches to the spatial infinity with $\delta h_E^{bc}=[h_E^{bc}]$ and $\delta \Psi_E^{bc}=[\Psi_E^{bc}]$.

    %Among others, below we shall demonstrate that the aforementioned criterion is satisfied by the four dimensional Einstein's gravity in both asymptotically flat and asymptotically AdS spacetimes as well as its higher derivative correction. 
Next let us demonstrate that the above criterion is satisfied by Einstein's gravity, whose Lagrangian form reads 
    \begin{equation}\label{EH}
        \mathbf{L}_{eh}=\frac{1}{16\pi}\bm{\epsilon}(R+\frac{6}{l^2}).
    \end{equation}
    Whence it is easy to show $\psi_{abcd}=\frac{1}{32\pi}(g_{ac}g_{bd}-g_{ad}g_{bc})$ and $\Psi_{ab}=\frac{1}{32\pi}h_{ab}$. Accordingly, we obtain
    \begin{eqnarray}
        \mathbf{B}=\frac{K}{8\pi}\hat{\bm{\epsilon}},\quad \mathbf{C}=-\frac{1}{16\pi}\dbar{A}\cdot\hat{\bm{\epsilon}},
         \quad \mathbf{F}=-\frac{1}{2}T_{bc}\delta 
        h^{bc}\hat{\bm\epsilon},
       \quad q_\xi^a=T^a{}_c\xi^c,
    \end{eqnarray}
    where $T_{bc}= -\frac{1}{8\pi}(K_{bc}-Kh_{bc})$ is the familiar Brown-York boundary energy-momentum tensor. 
    On the other hand, the corresponding Kerr-AdS black hole solution reads
 %\begin{eqnarray}
       % ds^2&=&-W(1+\frac{r^2}{l^2})dt^2+
       % \frac{2m}{U}(Wdt-\frac{a\sin^2\theta d\phi}{\Xi})^2+\frac{U}{V-2m}dr^2+\frac{r^2+a^2\cos^2\theta}{W\Xi}d\theta^2+
       % \frac{r^2+a^2}{\Xi}\sin^2\theta d\phi^2
    %\end{eqnarray}
    \begin{eqnarray}
        ds^2&=&-W(1+\frac{r^2}{l^2})dt^2+
        \frac{2m}{U}(Wdt-\frac{a\sin^2\theta d\phi}{\Xi})^2+\frac{U}{V-2m}dr^2+\frac{rU}{W\Xi}d\theta^2+
        \frac{r^2+a^2}{\Xi}\sin^2\theta d\phi^2
    \end{eqnarray}
    in the Boyer-Lindquist coordinates, 
    where 
    \begin{eqnarray}
        W=\frac{\sin^2\theta}{\Xi}+\cos^2\theta, \quad \Xi=1-\frac{a^2}{l^2},\quad U=\frac{r^2+a^2\cos^2\theta}{r},\quad V=\frac{1}{r}(1+\frac{r^2}{l^2})(r^2+a^2).
    \end{eqnarray}
    As alluded to in the beginning of this section, the coordinate system we are taking here is different from that in \cite{HB}. The main advantage of this coordinate system is that the corresponding Killing vector field $\frac{\partial}{\partial t}$ is non-rotating at infinity. With the Killing vector field normal to the black hole event horizon taking the form $\xi=\frac{\partial}{\partial t}+(\frac{a\Xi}{r_+^2+a^2}+\frac{a}{l^2} )\frac{\partial}{\partial\phi}$, we can obtain the black hole temperature and angular velocity as follows
    \begin{equation}\label{TO}
    T=\frac{r_+(1+\frac{a^2}{l^2}+3\frac{r_+^2}{l^2}-\frac{a^2}{r_+^2})}{4\pi(r_+^2+a^2)},\quad \Omega=\frac{a(1+\frac{r_+^2}{l^2})}{r_+^2+a^2},
\end{equation}
where $r_+$ denotes the black hole event horizon radius, satisfying $V(r_+)-2m=0$. 

To proceed, we shall choose the timelike boundary $\Gamma$ to be the surface of $r=\bar{r}$ and make the following coordinate transformation 
\begin{equation}\label{shift1}
    t\rightarrow \frac{\sqrt{V(\bar{r})-2m}}{\sqrt{V(\bar{r})}}t
\end{equation}
for the solution with $m=0$, which is actually the pure AdS solution in the Boyer-Lindquist coordinates and will serve as our reference spacetime.
After such a rescaling of the time coordinate, the metric of our reference spacetime reads
% \begin{eqnarray}
        %ds^2&=&-W\frac{V(\bar{r})-2m}{V(\bar{r})}(1+\frac{r^2}{l^2})dt^2+\frac{U}{V}dr^2+\frac{r^2+a^2\cos^2\theta}{W\Xi}d\theta^2+
        %\frac{r^2+a^2}{\Xi}\sin^2\theta d\phi^2,
   % \end{eqnarray}
     \begin{eqnarray}
        ds^2&=&-W\frac{V(\bar{r})-2m}{V(\bar{r})}(1+\frac{r^2}{l^2})dt^2+\frac{U}{V}dr^2+\frac{rU}{W\Xi}d\theta^2+
        \frac{r^2+a^2}{\Xi}\sin^2\theta d\phi^2,
    \end{eqnarray}
whereby one can choose 
 \begin{eqnarray}
       &&e^0\equiv \sqrt{W\frac{V(\bar{r})-2m}{V(\bar{r})}(1+\frac{r^2}{l^2})}dt ,\quad e^1\equiv \sqrt{\frac{U}{V}}dr, \nonumber\\
       &&e^2\equiv \sqrt{\frac{r^2+a^2\cos^2\theta}{W\Xi}}d\theta, \quad e^3\equiv\sqrt{\frac{r^2+a^2}{\Xi}}\sin\theta d\phi
    \end{eqnarray}
  as its orthonormal basis. Accordingly,  the non-vanishing basis components of the induced metric and extrinsic curvature at $\Gamma$ can be written as 
\begin{eqnarray}
&& h_{00}=-1+\frac{2ma^2\sin^2\theta}{\Xi\bar{r}^3
}+\mathcal{O}(\frac{1}{\bar{r}^5}), \quad h_{03}=-\frac{2mal\sin\theta\sqrt{W}}{\sqrt{\Xi}\bar{r}^3}+\mathcal{O}(\frac{1}{\bar{r}^5}), \quad h_{22}=1, \quad h_{33}=1+\frac{2ma^2\sin^2\theta}{\Xi\bar{r}^3}+\mathcal{O}(\frac{1}{\bar{r}^5}),\nonumber\\
&&K_{00}=-\frac{1}{l}+\frac{1}{2l\bar{r}^2}(l^2-{a^2\sin^2\theta})+\frac{m[-2l^2+{a^2}(1+\cos^2\theta)]}{l\Xi\bar{r}^3}+\mathcal{O}(\frac{1}{\bar{r}^4}), \quad
 K_{03}=\frac{ma\sin\theta\sqrt{W}}{\sqrt{\Xi} \bar{r}^3}+\mathcal{O}(\frac{1}{\bar{r}^4}),\nonumber\\
&& K_{22}=\frac{1}{l}+\frac{a^2+l^2-3a^2\cos^2\theta}{2l\bar{r}^2}-\frac{ml}{\bar{r}^3}+\mathcal{O}(\frac{1}{\bar{r}^4}),\quad K_{33}=\frac{1}{l}+\frac{l^2-a^2(1+\cos^2\theta)}{2l\bar{r}^2}-\frac{mlW}{\bar{r}^3}+\mathcal{O}(\frac{1}{\bar{r}^4}),
  \end{eqnarray}
whereby we can further obtain
\begin{equation}
    K=\frac{3}{l}+\frac{a^2+l^2-5a^2\cos^2\theta}{2l\bar{r}^2}+\frac{l^4-2a^2 l^2(1+3\cos^2\theta)+a^4(1-6\cos^2\theta+21\cos^4\theta)}{8l\bar{r}^4}-\frac{ml(l^2+a^2\sin^2\theta)}{\bar{r}^5}+\mathcal{O}(\frac{1}{\bar{r}^6}),
\end{equation} 
and the non-vanishing basis components of the  Brown-York tensor of the Kerr metric 
\begin{eqnarray}
&&T_{00}=-\frac{1}{4\pi l}-\frac{l^2-2a^2\cos^2\theta}{8\pi l\bar{r}^2}+\frac{ml[2+\frac{a^2}{l^2}(5-7\cos^2\theta)]}{8\pi\Xi\bar{r}^3}+\mathcal{O}(\frac{1}{\bar{r}^4}),\quad T_{03}=\frac{7ma\sin\theta\sqrt{W}}{8\sqrt{\Xi}\pi \bar{r}^3}+\mathcal{O}(\frac{1}{\bar{r}^4}),\nonumber\\
    && T_{22}=\frac{1}{4\pi l}-\frac{a^2\cos^2\theta}{8\pi l\bar{r}^2}+\frac{ml}{8\pi \bar{r}^3}+\mathcal{O}(\frac{1}{\bar{r}^4}),\quad 
    T_{33}=\frac{1}{4\pi l}-\frac{a^2(2\cos^2\theta-1)}{8\pi l\bar{r}^2}+\frac{ml[1+\frac{a^2}{l^2}(6-7\cos^2\theta)]}{8\pi\Xi \bar{r}^3}+\mathcal{O}(\frac{1}{\bar{r}^4})
\end{eqnarray}
with $m=0$ corresponding to the result for the reference  spacetime.  Note that $\sqrt{|h|}=\frac{\sqrt{(V-2m)U}r\sin\theta}{\Xi}$ is the same for the Kerr-AdS black hole and reference spacetime with $\frac{\bar{r}^3\sin\theta}{l\Xi} $ as the leading order term, but the induced metric of the Kerr-AdS black hole at $\Gamma$
is different from that of the reference spacetime. Luckily, this does not destroy the applicability of the background subtraction method according to our criterion. 
%i.e., 
%\begin{equation}\label{bcanother}
%    [h_{00}]=\mathcal{O}(\frac{1}{\bar{r}^3})
%\quad [h_{03}]=\mathcal{O}(\frac{1}{\bar{r}^3}),\quad [h_{33}]=\mathcal{O}(\frac{1}{\bar{r}^3}).
%\end{equation}
To be more specific, we have the induced metrics
\begin{eqnarray}
 h'_{00}&=&-1-\frac{2a\delta a \sin^2\theta}{l^2\Xi^2 W}+\frac{1}{\bar{r}^3}[\frac{2ma^2\sin^2\theta+2W\Xi l^2\delta m}{\Xi}+\frac{4ma\delta a\sin^2\theta(l^2- a^2\cos 2\theta)}{Wl^2\Xi^2}]+\mathcal{O}(\frac{1}{\bar{r}^4}), \nonumber\\
 h'_{03}&=&-\frac{1}{\bar{r}^3}[\frac{2(m+\delta m)al\sin\theta\sqrt{W}}{\sqrt{\Xi}}+\frac{2m\delta a\sin^2\theta(l^4+3a^2l^2\sin^2\theta-a^4\cos^2\theta)}{l^3\Xi^2\sin\theta\sqrt{W\Xi}}]+\mathcal{O}(\frac{1}{\bar{r}^4}),\nonumber\\
 h'_{22}&=&1+\frac{2a\delta a \cos^2\theta}{l^2W\Xi}+\frac{2a\delta a\cos^2\theta}{\bar{r}^2}+\mathcal{O}(\frac{1}{\bar{r}^4}),\nonumber\\
  h'_{33}&=&1+\frac{2a\delta a }{l^2\Xi}+\frac{2a\delta a}{ \bar{r}^2}+\frac{1}{\bar{r}^3}[\frac{2(m+\delta m)a^2\sin^2\theta}{\Xi}+\frac{4ma\delta a(l^2+a^2)\sin^2\theta}{l^2\Xi^2}]+\mathcal{O}(\frac{1}{\bar{r}^4}),\nonumber\\
  h'^0_{00}&=&-1-\frac{ 2a \delta a\sin^2\theta}{l^2\Xi^2W}+\frac{2\delta m l^2}{\bar{r}^3}+\mathcal{O}(\frac{1}{\bar{r}^4}), \quad
   h'^0_{03}=0, \nonumber\\
   h'^0_{22}&=&1+\frac{2a\delta a\cos^2\theta}{l^2W\Xi}+\frac{2a\delta a\cos^2\theta}{\bar{r}^2}+\mathcal{O}(\frac{1}{\bar{r}^4}),\quad
  h'^0_{33}=1+\frac{2a\delta a }{l^2\Xi }+\frac{2a\delta a}{ \bar{r}^2}+\mathcal{O}(\frac{1}{\bar{r}^4})
\end{eqnarray}
for the first order perturbed Kerr-AdS black hole and  reference spacetime. By $\delta h^{ab}=-h^{ac}h^{bd}\delta h_{cd}$, one can further obtain
\begin{eqnarray}
 \delta{h'^{00}}&=&\frac{2a\delta a\sin^2\theta}{l^2\Xi^2W}-\frac{1}{\bar{r}^3}[{2\delta m Wl^2}+\frac{4ma\delta a\sin^2\theta}{W\Xi^2}]+\mathcal{O}(\frac{1}{\bar{r}^4}), \nonumber\\
\delta{h'^{03}}&=&\frac{1}{\bar{r}^3}[\frac{2a\delta ml \sin\theta\sqrt{W}}{\sqrt{\Xi}}+\frac{2m\delta a\sin\theta [l^4\Xi-3a^4\cos^2\theta+a^2l^2(7-5\cos^2\theta)]}{l^3\Xi\sqrt{W\Xi} }]+\mathcal{O}(\frac{1}{\bar{r}^4}), \nonumber\\
\delta{h'^{22}}&=&-\frac{2a\delta a \cos^2\theta}{l^2W\Xi}-\frac{2a\delta a\cos^2\theta }{\bar{r}^2}+\mathcal{O}(\frac{1}{\bar{r}^4}), \nonumber\\
\delta{h'^{33}}&=&-\frac{2a\delta a}{l^2\Xi}-\frac{2a\delta a}{\bar{r}^2}-\frac{1}{\bar{r}^3}[\frac{2a^2\delta m \sin^2\theta}{\Xi}+\frac{4ma\delta a\sin^2\theta}{\Xi}]+\mathcal{O}(\frac{1}{\bar{r}^4}),\nonumber\\
 \delta{h'^{0}}^{00}&=&\frac{2a\delta a\sin^2\theta}{l^2\Xi^2W}-\frac{2\delta m l^2}{\bar{r}^3}+\mathcal{O}(\frac{1}{\bar{r}^4}), \quad
\delta{h'^{0}}^{03}=0, \nonumber\\
\delta{h'^{0}}^{22}&=&-\frac{2a\delta a\cos^2\theta}{l^2W\Xi}-\frac{2a\delta a\cos^2\theta}{\bar{r}^2}+\mathcal{O}(\frac{1}{\bar{r}^4}), \quad
\delta{h'^{0}}^{33}=-\frac{2a\delta a}{l^2\X}-\frac{2a\delta a}{\bar{r}^2}+\mathcal{O}(\frac{1}{\bar{r}^4}).
\end{eqnarray}
Then it follows from a straightforward computation that $\int_{\mathcal{S}}\xi\cdot[\mathbf{F}]$ is proportional to $\int_{-1}^1dx(1-3x^2)=0$ with $x=\cos\theta$, which guarantees the validity of the first law of black hole thermodynamics. On the other hand, we have 
\begin{eqnarray}\label{pure}
T^t{}_t=\frac{1}{4\pi l}+\frac{l^2-2a^2\cos^2\theta}{8\pi l \bar{r}^2 }-\frac{ml[2+\frac{a^2}{l^2}(1-3\cos^2\theta)]}{8\pi\Xi \bar{r}^3}+\mathcal{O}(\frac{1}{\bar{r}^4}),\quad 
T^t{}_\phi=\frac{3mal\sin^2\theta}{8\pi \Xi \bar{r}^3}+\mathcal{O}(\frac{1}{\bar{r}^4})
\end{eqnarray}
with $m=0$ corresponding to the result for the reference spacetime, whereby the finiteness of $[H_\xi]$ is apparent. In particular, we have the familiar ADM mass and angular momentum as follows
\begin{equation}
    M\equiv [H_{\frac{\partial}{\partial t}}]=\frac{m}{\Xi^2},\quad J\equiv-[H_{\frac{\partial}{\partial\phi}}]=Ma,
\end{equation}
which is finite. Thus the background subtraction method is applicable to Einstein's gravity in asymptotically AdS spacetime.

Finally, as detailed in \cite{HB}, the criterion for the applicability of the background subtraction method is also satisfied by the higher derivative gravity as the correction to Einstein's gravity within the framework of effective field theory. Moreover, the corresponding Gibbs free energy receives the contribution only from the bulk term, as the same as that for Einstein's gravity. However, there are two additional effects from the higher derivative terms. First, suppose that the pure AdS is still the solution to the aforementioned higher derivative gravity theory and $\lambda$ is a small constant, then Eq. (\ref{variationradius}) evaluated on such a pure AdS solution with
 $\delta g_{ab}=\lambda g_{ab}$ gives rise to 
 \begin{equation}
     \delta \mathbf{L}=0,
 \end{equation}
 where we have used $d\Theta=0$ due to the fact that the pure AdS is a space of constant curvature. Note that both the traceless part of Ricci tensor $\mathcal{R}_{ab}$ and Weyl tensor $C_{abcd}$ vanish for the pure AdS, thus the above equation implies that the four dimensional AdS radius will get corrected only by the $\frac{\varepsilon}{16\pi}\bm{\epsilon}R^n$ term with $n>2$ as follows
\begin{equation}\label{adsradius}
    R+\frac{12}{l^2}+\varepsilon(2-n)R^n=0.
\end{equation}
Second, the expression for the ADM mass and angular momentum will get corrected 
only by the higher derivative terms $\frac{\varepsilon}{16\pi}\bm{\epsilon} R^n$ and $\frac{\varepsilon}{16\pi}\bm{\epsilon}R^nC^2$
with $C^2\equiv C_{abcd}C^{abcd}$. It turns out that 
 the corrections to the ADM mass and angular momentum are equally proportional to the expression for Einstein's gravity with 
 the proportional constant $\varepsilon nR^{n-1}$ for the former higher derivative correction and $\frac{\varepsilon 4R^n}{l_e^2}$ for the latter one, where $l_e$ denotes the corrected AdS radius.

\iffalse\fi

%Then we further have
%\begin{equation}
%\nabla^eC_{deaf}=\mathcal{O}(\frac{1}{\bar{r}^4}), \quad h_a{}^eh_c{}^fC_{efdb}n^d=\mathcal{O}(\frac{1}{\bar{r}^6}),\quad \nabla_{[e}C_{bc]af}=\mathcal{O}(\frac{1}{\bar{r}^4}),\quad h_a{}^eh_c{}^f\tilde{C}_{efdb}n^d)=\mathcal{O}(\frac{1}{\bar{r}^5}).
%\end{equation}

%Here is a remark. For Einstein gravity, there is no contribution from the bulk term in the asymptotically flat spacetime and no boundary contribution in the asympotically AdS spacetime.

%Two exmaples

\section{Higher derivative corrections to Kerr-AdS black hole thermodynamics}\label{33}

With the preparation made in the previous section, we shall apply the background subtraction method to calculate the first order corrections to Kerr-AdS black hole thermodynamics by the higher derivative terms. To be more precise, we shall consider the Einstein-Hilbert term (\ref{EH}) corrected by the following high derivative terms\footnote{Here we have ignored all the other quadratic and cubic terms because their corrections vanish due to the fact that the involved Kerr-AdS metric has $\mathcal{R}_{ab}=0$ as well as an orientation reversing isometry $\theta\rightarrow \pi-\theta$. } 
\begin{equation}
  \mathbf{L}_{hd} = \frac{1}{16\pi}\bm{\epsilon}(\varepsilon_1L^2R^2+\varepsilon_2L^2C^2+\varepsilon_3L^4R^3+\varepsilon_4L^4RC^2+\varepsilon_5L^4C_{abcd}C^{cdef}C^{ab}{}_{ef}),
\end{equation}
where $L$ as some UV length scale has been inserted to ensure $\varepsilon_i$ dimensionless. 
First, according to Eq. (\ref{adsradius}), the AdS radius gets shifted as follows
\begin{equation}
    \frac{12}{l_e^2}=\frac{12}{l^2}+\varepsilon_3L^4(\frac{12}{l^2})^3=-R.
\end{equation}
Then we follow the trick devised in \cite{Xiao2} to decompose the full bulk action  as follows
\begin{equation}
  I'= \int_M( \mathbf{L}'_{eh}+\mathbf{L}'_{hd})
\end{equation}
with 
\begin{equation}
    \mathbf{L}'_{eh}=\frac{1}{16\pi}\bm{\epsilon}(R+\frac{6}{l_e^2}), \quad \mathbf{L}'_{hd}=\frac{1}{16\pi}\bm{\epsilon}(\frac{6}{l^2}-\frac{6}{l_e^2})+\mathbf{L}_{hd}.
\end{equation}
In light of the analysis detailed in the previous section, the first order corrected Gibbs free energy can be obtained by evaluating the corresponding bulk $[I'_E]$ on the Kerr-AdS metric and pure AdS metric with $l_e$ as the AdS radius. The underlying reason is the same as documented in \cite{RS} for the asymptotically flat case and can be recapitulated as follows.
%\footnote{Such a strategy for the first order correction calculation without resorting to the full perturbed first order solution has stimulated a lot of follow-up studies\cite{Cheung,Cremonini,Melo,Bobev,Xiao1,Cassani,Noumi,Ma,Zatti,Massai,Hu,Xiao2,Ma2}.}. 
First, as pointed out at the end of the previous section, there is no contribution to the Gibbs free energy from the associated boundary terms. In addition, according to Eq. (\ref{variationradius}) and Eq. (\ref{boundaryex}),  the contribution to the Gibbs free energy from $\mathbf{L}'_{eh}$ evaluated on the full first order corrected solution is the same as that evaluated on the aforementioned Kerr-AdS metric, because their difference comes solely from the boundary term $\int_\infty [\mathbf{F}_E]=\int_\infty d\tau\wedge\frac{\partial}{\partial\tau}\cdot \mathbf{F}_E=\beta\int_{\mathcal{S}}\xi\cdot[\mathbf{F}]=0$ with $\mathbf{F}_E=-i\mathbf{F}$. Last but not least, similarly, the contribution to the first order corrected Gibbs free energy from $\mathbf{L}'_{hd}$ evaluated at the full first order corrected solution is also the same as that evaluated on the aforementioned Kerr-AdS metric because their difference, determined by the first term in Eq. (\ref{variationradius}), starts apparently from the second order of $\varepsilon_i$, where one $\varepsilon_i$ comes from the coefficients in $\mathbf{L}'_{hd}$ and the other $\varepsilon_i$ comes from the variation of the metric induced by the higher derivative terms.

With this in mind, we first calculate the bulk contribution to $[I'_E]$ from $\mathbf{L}'_{eh}$. As such, we have 
\begin{equation}
    \int_{BH} \mathbf{L}'_{ehE}  =\frac{\beta}{16\pi} \int_{0}^{2\pi} d\phi\int_{0}^{\pi} d\theta \int_{r_+}^{\bar{r}}dr \frac{6}{l_e^2}\sqrt{|g|}=\frac{\beta (\bar{r}-r_+)(\bar{r}^2+r_+\bar{r}+r_+^2+a^2)}{2l_e^2\Xi}
\end{equation}
and  
\begin{equation}
   \int_{AdS} \mathbf{L}'_{ehE}=\frac{\beta}{16\pi} \int_{0}^{2\pi} d\phi\int_{0}^{\pi} d\theta \int_{0}^{\bar{r}}dr \frac{6}{l_e^2}\sqrt{|g|}
   %=\frac{\beta (\bar{r}^2+a^2)(\bar{r}^3-ml_e^2)}{2l_e^2\bar{r}^2\Xi}
   =\frac{\beta}{2l_e^2\Xi}(\bar r^3+{a^2 \bar r}-m l_e^2)+\mathcal O(\frac{1}{\bar{r}}), 
\end{equation}
where $\sqrt{|g|}_{BH}=\frac{rU\sin\theta}{\Xi}$ and $\sqrt{|g|}_{AdS}=\frac{\sqrt{V(\bar{r})-2m}}{\sqrt{V(\bar{r})}}\frac{rU\sin\theta}{\Xi}$ are used. Then with the background subtraction method, the resulting contribution to $[I'_E]$ is obtained as
\begin{equation}
-\frac{\beta(r_+^3+a^2r_+-ml_e^2)}{2l_e^2\Xi}
\end{equation}
by taking the limit $\bar{r}\rightarrow\infty$.
Next by the same token and with 
\begin{eqnarray}
    C^2&=&\frac{48m^2(r^6-15a^2r^4x^2+15a^4r^2x^4-a^6x^6)}{(r^2+a^2x^2)^6},\nonumber\\
    C_{abcd}C^{cdef}{C^{ab}}_{ef}&=&\frac{96m^3r(r^8-36a^2r^6x^2+126a^4r^4x^4-84a^6r^2x^6+9a^8x^8)}{(r^2+a^2x^2)^9}
\end{eqnarray}
for the Kerr-AdS metric, one can further obtain the corresponding contribution to $[I'_E]$ from $\mathbf{L}'_{hd}$. Accordingly, we end up with the following first order corrected Gibbs free energy  
\begin{eqnarray}
    G&\simeq&-\frac{r_+^3+a^2r_+-ml_e^2}{2l_e^2\Xi}+(12\varepsilon_1\frac{L^2}{l_e^2}-216\varepsilon_3\frac{L^4}{l_e^4})\frac{r_+^3+a^2r_+-ml_e^2}{l_e^2\Xi}-(4\varepsilon_2-48\varepsilon_4\frac{L^2}{l_e^2})\frac{L^2m^2r_+(r_+^2-a^2)}{\Xi(r_+^2+a^2)^3}-\nonumber\\
    &&4\varepsilon_5\frac{L^4m^3(7r_+^6-35a^2r_+^4+21a^4r_+^2-a^6)}{7\Xi(r_+^2+a^2)^6},
\end{eqnarray}
where $\simeq$ means that the equation holds at the first order of $\varepsilon_i$.

%\begin{equation}
%  \int_{-1}^1dx \int_{r^+}^{\infty} dr C^{abcd}C_{abcd}\sqrt{-g}=\frac{32m^2r^+({r^+}^2-a^2)}{\Xi({r^+}^2+a^2)^3}
%\end{equation}
%\begin{equation}
 %   C_{abcd}C^{cdef}{C^{ab}}_{ef}=-\frac{96m^3r(r^8-36a^2r^6x^2+126a^4r^4x^4-84a^6r^2x^6+9a^8x^8)}{(r^2+a^2x^2)^9}
%\end{equation}
%\begin{equation}
%   \int_{-1}^1dx\int_{r^+}^{\infty}dr C^{abcd}C_{cdmn}{C_{ab}}^{mn}\sqrt{-g}=-\frac{32m^3(7{r^+}^6-35a^2{r^+}^4+21a^4{r^+}^2-a^6)}{7\Xi({r^+}^2+a^2)^6}
%\end{equation}
%\begin{equation}
%  \lim_{R\to \infty} \beta\int_{-1}^1dx \int_{r^+}^{R} dr \sqrt{-g}-\beta(1-\frac{ml^2}{R^3})\int_{-1}^1dx \int_{0}^{R} dr \sqrt{-g}=-\beta\frac{2{r^+}^3+a^2r^+-ml^2}{3\Xi}
%\end{equation}
To obtain the corresponding higher derivative corrections to other thermodynamic quantities from the above Gibbs free energy in the grand canonical ensemble, we should view $m$ as well as $r_+$ and $a$ as functions of $T$ and $\Omega$ implicitly through Eq.  (\ref{TO}), whereby one can get the following partial derivatives 
\begin{eqnarray}
   &&\frac{\partial r_+}{\partial T}=
    \frac{4\pi r_+^2 l_e^2(r_+^2-a^2)}{4r_+^4-(r_+^2+l_e^2)(r_+^2+a^2)},\quad
    \frac{\partial r_+}{\partial \Omega}=
    \frac{4al_e^2r_+^3 }{4r_+^4-(r_+^2+l_e^2)(r_+^2+a^2)},\nonumber\\
    &&\frac{\partial a}{\partial T}=
    \frac{8\pi al_e^4\Xi r_+^3 }{[4r_+^4-(r_+^2+l_e^2)(r_+^2+a^2)](r_+^2+l_e^2)}
,\quad
    \frac{\partial a}{\partial \Omega}=
    \frac{3r_+^6+(8a^2-l_e^2)r_+^4+a^2(a^2+4l_e^2)r_+^2+a^4l_e^2 }{[4r_+^4-(r_+^2+l_e^2)(r_+^2+a^2)](1+\frac{r_+^2}{l_e^2})}.
\end{eqnarray}
Then the corresponding first order corrected entropy, angular momentum, and mass can be calculated out as follows
\begin{eqnarray}
    S&=&-\frac{\partial G}{\partial T}\simeq\frac{\pi (r_+^2+a^2)}{\Xi}-\varepsilon_1\frac{24\pi L^2(r_+^2+a^2)}{l_e^2\Xi}+\varepsilon_2\frac{4\pi L^2(r_+^2+l_e^2)}{l_e^2\Xi}+\varepsilon_3\frac{432\pi L^4(r_+^2+a^2)}{l_e^4\Xi}-\varepsilon_4\frac{48\pi L^4(r_+^2+l_e^2)}{l_e^4\Xi}\nonumber\\
   % &=&\varepsilon_5 \frac{2\pi L^4(r_+^2+l_e^2)^2[-21 r_+^{10}+7(3l^2-2a^2)r_+^8+2a^2(98a^2-7l_e^2)r_+^6-2a^4(33a^2+28l_e^2)r_+^4+a^6(a^2-18l_e^2)r_+^2+3a^8l_e^2]}{7l_e^4\Xi r_+^2(r_+^2+a^2)^3[4r_+^4-(r_+^2+a^2)(r_+^2+l^2)]}\\
    &&-\varepsilon_5 \frac{8\pi L^4m^2[a^8(3l_e^2+r_+^2)-6a^6r_+^2(3l_e^2+11r_+^4)-28a^4r_+^4(2l_e^2-7r_+^2)-14a^2r_+^6(l_e^2+r_+^2)+21r_+^8(l_e^2-r_+^2)]}{7\Xi(r_+^2+a^2)^5[4r_+^4-(r_+^2+a^2)(r_+^2+l_e^2)]},\label{entropy}\\
    J&=&-\frac{\partial G}{\partial \Omega}\simeq\frac{ma}{\Xi^2}-\varepsilon_1\frac{24maL^2}{l_e^2\Xi^2}+\varepsilon_2\frac{4maL^2}{l_e^2\Xi^2}+\varepsilon_3\frac{432maL^4}{l_e^4\Xi^2}-\varepsilon_4\frac{48ma L^4}{l_e^4\Xi^2}\nonumber\\
    %&=&-\varepsilon_5 \frac{(r_+^2+l_e^2)^2L^4[21ar_+^{12}+7a(11a^2-19l_e^2)r_+^{10}+7a(-40a^5+17a^2l_e^2+2l_e^4)r_+^8+6a^3(-6a^4-5a^2l_e^2+7l_e^4)r_+^6+a^5(43a^4+10a^2l_e^2+10l_e^4)r_+^4+a^{7}(-a^{4}+35a^2l_e^2-18l_e^4)r_+^2-a^11l_e^2]}{7l_e^6\Xi^2 r_+^3(r_+^2+a^2)^3[4r_+^4-(r_+^2+a^2)(r_+^2+l_e^2)]}\nonumber\\
   % &=&-\varepsilon_5 \frac{4m^2L^4[-a^{11}(r_+^2+l_e^2)+a^9r_+^2(35l_e^2+43r_+^2)-2a^7r_+^2(9l_e^4-5l_e^2r_+^2+18r_+^4)+10a^5r_+^4(l_e^4-3l_e^2r_+^2-28r_+^4)+7a^3r_+^6(6l_e^4+17l_e^2r_+^2+11r_+^4)+7ar_+^8(2l_e^4-19l_e^2r_+^2+3r_+^4)]}{7l_e^2\Xi^2 r_+(r_+^2+a^2)^5[4r_+^4-(r_+^2+a^2)(r_+^2+l_e^2)]}\nonumber\\
    &&+\varepsilon_5 \{\frac{4m^2L^4[-a^{11}(r_+^2+l_e^2)+a^9r_+^2(35l_e^2+43r_+^2)-2a^7r_+^2(9l_e^4-5l_e^2r_+^2+18r_+^4)+10a^5r_+^4(l_e^4-3l_e^2r_+^2-28r_+^4)]}{7l_e^2\Xi^2 r_+(r_+^2+a^2)^5[4r_+^4-(r_+^2+a^2)(r_+^2+l_e^2)]}\nonumber\\
&&+\frac{4m^2L^4[7a^3r_+^6(6l_e^4+17l_e^2r_+^2+11r_+^4)+7ar_+^8(2l_e^4-19l_e^2r_+^2+3r_+^4)]}{7l_e^2\Xi^2 r_+(r_+^2+a^2)^5[4r_+^4-(r_+^2+a^2)(r_+^2+l_e^2)]}\},\label{thermalangular}\\
    M&=&G+TS+\Omega J\simeq\frac{m}{\Xi^2}-\varepsilon_1\frac{24mL^2}{l_e^2\Xi^2}+\varepsilon_2\frac{4mL^2}{l_e^2\Xi^2}+\varepsilon_3\frac{432mL^4}{l_e^4\Xi^2}-\varepsilon_4\frac{48m L^4}{l_e^4\Xi^2}\nonumber\\
   %  &=&\varepsilon_5 \frac{4m^2L^4[2a^{10}l_e^2-a^8(l_e^4+36l_e^2r_+^2+85r_+^4)+4a^6r_+^2(5l_e^4+9l_e^2r_+^2+55r_+^4)-14a^4r_+^4(l_e^4+12l_e^2r_+^2-7r_+^4)+14a^2r_+^6(-2l_e^4+3l_e^2r_+^2+2r_+^4)+7r_+^8(l_e^4+4l_e^2r_+^2-3r_+^4)]}{7l_e^2\Xi^2 r_+(r_+^2+a^2)^5[4r_+^4-(r_+^2+a^2)(r_+^2+l_e^2)]}\nonumber\\
     &&-\varepsilon_5 \{\frac{4m^2L^4[2a^{10}l_e^2-a^8(l_e^4+36l_e^2r_+^2+85r_+^4)+4a^6r_+^2(5l_e^4+9l_e^2r_+^2+55r_+^4)-14a^4r_+^4(l_e^4+12l_e^2r_+^2-7r_+^4)]}{7l_e^2\Xi^2 r_+(r_+^2+a^2)^5[4r_+^4-(r_+^2+a^2)(r_+^2+l_e^2)]}\nonumber\\
&&+\frac{4m^2L^4[14a^2r_+^6(-2l_e^4+3l_e^2r_+^2+2r_+^4)+7r_+^8(l_e^4+4l_e^2r_+^2-3r_+^4)]}{7l_e^2\Xi^2 r_+(r_+^2+a^2)^5[4r_+^4-(r_+^2+a^2)(r_+^2+l_e^2)]}\}.\label{thermalmass}
\end{eqnarray}
By comparing the above corrected mass and angular momentum with those obtained through the ADM formula stated at the end of the previous section, one can know that the higher derivative terms except the one associated with  $\varepsilon_5$ do not correct the parameters $m$ and $a$ appearing in the Kerr-AdS metric under consideration. On the other hand, according to Eq. (\ref{entropyformula}), $\nabla_d\xi_b=\kappa \bar{\bm{\epsilon}}_{db}$ and $\bm{\epsilon}=\bar{\bm{\epsilon}}\wedge\tilde{\bm{\epsilon}}$ at the bifurcation surface $\mathcal{B}$ with $\bar{\bm{\epsilon}}$ the binormal to $\mathcal{B}$ and $\tilde{\bm{\epsilon}}$ the induced volume at $\mathcal{B}$, the black hole entropy can be expressed by the Wald formula as\cite{Wald,IW}
\begin{equation}
    S=-2\pi\int_\mathcal{B} \psi^{cadb}\bar{\bm\epsilon}_{db}\bar{\bm{\epsilon}}_{ca}\tilde{\bm{\epsilon}},
\end{equation}
whereby the first order corrected entropy can be written as follows
\begin{equation}\label{waldentropy}
    S\simeq \frac{A(\mathcal{B})}{4}+\frac{\delta A(\mathcal{B})}{4}-2\pi\int_\mathcal{B}\psi_{hd}^{cadb}\bar{\bm\epsilon}_{db}\bar{\bm{\epsilon}}_{ca}\tilde{\bm{\epsilon}}.
\end{equation}
Here $A(\mathcal{B})$ denotes the  horizon area of the Kerr-AdS black hole with the AdS radius $l_e$, given by 
\begin{equation}
A(\mathcal{B})=\int_{\mathcal{B}}\tilde{\bm{\epsilon}}=\frac{4\pi(r_+^2+a^2)}{\Xi},
\end{equation}
where we have used $\tilde{\bm{\epsilon}}=\frac{(r_+^2+a^2)\sin\theta}{\Xi}d\theta\wedge d\phi$. While $\delta A(\mathcal{B})$ represents the area correction due to the deviation of the full first order corrected black hole solution from the Kerr-AdS metric by the higher derivative terms. On the other hand, the third term in Eq. (\ref{waldentropy}), evaluated at the aforementioned Kerr-AdS black hole, gives rise to the rest corrections from the higher derivative terms. In particular, by $\bar{\bm\epsilon}=\frac{r_+^2+a^2\cos^2\theta}{r_+^2+a^2}dt\wedge dr$ and
\begin{eqnarray}
\psi_{hd}^{abcd}&=&\frac{1}{16\pi}\{(2\varepsilon _1L^2R+3\varepsilon _3L^4R^2+\varepsilon_4L^4C^2+\varepsilon_5L^4C^2)g^{a[c}g^{d]b}+(2\varepsilon_2L^2+2\varepsilon_4L^4R)C^{abcd}+\nonumber\\
    &&3\varepsilon_5L^4[ C^{abmn}{C_{mn}}^{cd}-(C^{mnpa}{C_{mnp}}^{[c}g^{d]b}-C^{mnpb}{C_{mnp}}^{[c}g^{d]a})]\},
\end{eqnarray}
we can calculate it out explicitly as follows
\begin{eqnarray}\label{WS}
    -2\pi\int_\mathcal{B}\psi_{hd}^{cadb}\bar{\bm\epsilon}_{db}\bar{\bm{\epsilon}}_{ca}\tilde{\bm{\epsilon}}
&=&-\varepsilon_1\frac{24L^2\pi(r_+^2+a^2)}{l_e^2\Xi}+\varepsilon_2\frac{4\pi L^2(r_+^2+l_e^2)}{l_e^2\Xi}+\varepsilon_3\frac{432\pi L^4(r_+^2+a^2)}{l_e^4\Xi}-\varepsilon_4\frac{48\pi L^4(r_+^2+l_e^2)}{l_e^4\Xi}\nonumber\\
    &&+\varepsilon_4\frac{48 \pi L^4 m^2(a^4-10a^2 r_+^2+5r_+^4)}{5\Xi (r_+^2+a^2)^4}+\varepsilon_5 \frac{ 24\pi L^4 m^2(a^4-10a^2 r_+^2+5r_+^4)}{5\Xi (r_+^2+a^2)^4} .
\end{eqnarray}
By inspection of Eq. (\ref{entropy}) and (\ref{WS}), we find that the higher derivative terms associated with $\varepsilon_1$, $\varepsilon_2$ and $\varepsilon_3$ do not induce the additional variation of the black hole horizon area while the terms associated with $\varepsilon_4$ and $\varepsilon_5$ do so.

As a simple check of the validity of our result, we like to conclude this section by applying it to the corrections induced  by the Gauss-Bonnet term
\begin{equation}
    R^2-4R_{ab}R^{ab}+R_{abcd}R^{abcd}=\frac{1}{6}R^2-2\mathcal{R}_{ab}\mathcal{R}^{ab}+C^2,
\end{equation}
which is well known to be a topological term in four dimension and does not correct the solution to Einstein's gravity.  On the one hand, it follows from the remark made at the end of the previous section that the correction to the ADM mass and angular momentum from the $R^2$ term is precisely canceled out by that from the $C^2$ term. Thus we end up with the ADM mass and angular momentum uncorrected, which is consistent with Eq. (\ref{thermalangular}) and Eq. (\ref{thermalmass}). On the other hand, note that $\mathcal{R}_{ab}=0$ when evaluated on the Kerr-AdS black hole, so the corresponding entropy correction via Eq. (\ref{WS}) turns out to be exactly the same as that via Eq. (\ref{entropy}) as it should be the case.

\section{Conclusion}
With the viable background subtraction method, we have accomplished the calculation for the first order corrected Gibbs free energy as well as the corresponding entropy, angular momentum, and mass in the grand canonical ensemble by the higher derivative terms up to the cubic of Riemann tensor.
To the best of our knowledge, this is a new result, which otherwise would be difficult to obtain by the much more involved covariant counterterm method.
The validity of our result is further substantiated by examining the corrections induced by the Gauss-Bonnet term. Although our strategy does not require us to solve the full first order corrected black hole solutions by the higher derivative terms, we can extract some generic information about the first order corrected black hole solutions by comparing our results obtained via the Euclidean approach with those obtained via the ADM and Wald formulas in Lorentzian signature. 

Among others, there are two interesting issues worthy of further investigation. One is the first order corrections to our Kerr-AdS black hole by the higher derivative terms than the cubic of Riemann tensor. The other is to apply the background subtraction method to calculate the higher derivative corrections to thermodynamic quantities for the higher dimensional rotating black holes, which turns out to be much more sophisticated using the covariant counterterm method.  We hope to report our investigation along this line somewhere else in the future. 

\begin{acknowledgments}
This work is partially supported by the National Key Research and Development Program of China with Grant No. 2021YFC2203001 as well as the National Natural Science Foundation of China with Grant Nos. 12075026 and 12361141825. 
%In addition, this work was done in part during the workshop "Holographic Duality and Models of Quantum Computation" held at Tsinghua Southeast Asia Center on Bali, Indonesia (2024). 
%HZ is also grateful to Pengju Hu, Hong Lv, Liang Ma, Yi Pang, and Yong Xiao for sharing their wisdom on higher derivative corrections. He would also like to thank Yu Tian and Amos Yarom for stimulating discussions on this work. 

\end{acknowledgments}


\begin{thebibliography}{1}
%\bibitem{Witten} E. Witten, arXiv: 2412.16795 [hep-th].
\bibitem{Hartle}J. B. Hartle and S. W. Hawking, Phys. Rev. D \textbf{13}, 2188 (1976).
\bibitem{GH}G. W. Gibbons and S. W. Hawking, Phys. Rev. D \textbf{15}, 2752 (1976).

\bibitem{HP}S. W. Hawking and D. N.  Page, Commun. Math. Phys. \textbf{87}, 577 (1983).
\bibitem{HH}S. W. Hawking and G. T. Horowitz, Class. Quant. Grav. \textbf{13}, 1487 (1996).

\bibitem{GPP}G. W. Gibbons, M. J. Perry, and C. N. Pope, Class. Quant. Grav. \textbf{22}, 1503 (2005).


\bibitem{BFS}M. Bianchi, D. Z. Freedman, and K. Skenderis, Nucl. Phys. B \textbf{631}, 159 (2002).
\bibitem{PS}I. Papadimitriou and K. Skenderis, JHEP \textbf{0508}, 004 (2005).
\bibitem{MM}R. B. Mann and D. Marolf, Class. Quant. Grav. \textbf{23}, 2927 (2006).

\bibitem{HB}W. Guo, X. Guo, X. Lan, H. Zhang, and W. Zhang, Phys. Rev. D \textbf{111}, 084088 (2025).






%\bibitem{DG}S. Dutta and R. Gopakumar, Phys. Rev. D 74, 044007(2006).


\bibitem{RS}H. S. Reall and J. E. Santos, JHEP \textbf{04}, 021 (2019).



\bibitem{Cheung}C. Cheung, J. Liu, and G. N. Remmen, Phys. Rev. D \textbf{100}, 046003 (2019).

\bibitem{Cremonini}S. Cremonini, C. R. T. Jones, J. T. Liu, and B. McPeak, JHEP \textbf{09}, 003 (2020).

\bibitem{Belgium}N. Bobev, A. M. Charles, K. Hiristov, and V. Reys, Phys. Rev. Lett. \textbf{125}, 131601 (2020).

\bibitem{Melo}J. F. Melo and  J. E. Santos, Phys. Rev. D \textbf{103}, 066008 (2021).
\bibitem{Bobev}N. Bobev, V. Dimitrov, V. Reys, and A. Vekemans, Phys. Rev. D \textbf{106}, L121903 (2022).

\bibitem{Xiao1}Y. Xiao, Phys. Rev. D \textbf{106}, 064041 (2022).
\bibitem{Cassani}D. Cassani, A. Ruiperez, and E. Turetta, JHEP \textbf{11}, 059 (2022).

\bibitem{Noumi}T. Noumi and H. Satake, JHEP \textbf{12}, 130 (2022).
\bibitem{Cassani1}D. Cassani, A. Ruiperez, and E. Turetta, JHEP \textbf{06}, 203 (2023).

\bibitem{Ma}L. Ma, Y. Pang, and H. Lu,  JHEP \textbf{06}, 087 (2023).
\bibitem{Zatti}M. Zatti, JHEP \textbf{11}, 185 (2023).
\bibitem{Hu}P. Hu, L. Ma, H. Lu, and Y. Pang, Sci. China Phys. Mech. Astron. \textbf{67}, 280412 (2024).



\bibitem{Ma2}L. Ma, P. Hu, Y. Pang, and H. Lu, Phys. Rev. D \textbf{110}, L021901 (2024).

\bibitem{Cano}P. A. Cano and M. David, JHEP \textbf{03}, 036 (2024).

\bibitem{Massai}S. Massai, A. Ruiperez, and M. Zatti, JHEP \textbf{04}, 150 (2024).

\bibitem{Cassani2}D. Cassani, A. Ruiperez, and E. Turetta, JHEP \textbf{05}, 276 (2024).


\bibitem{Xiao2}Y. Xiao and Y. Liu, Phys. Rev. D \textbf{110}, 104043 (2024).


\bibitem{Wald}
R. M. Wald, Phys. Rev. D \textbf{48}, R3427 (1993).
\bibitem{IW}
V. Iyer and R. M. Wald, Phys. Rev. D \textbf{50}, 846 (1994).

%\bibitem{Iyer}
%V.~Iyer and R.~M.~Wald,
%``A Comparison of Noether charge and Euclidean methods for computing the entropy of stationary black holes,''
%Phys. Rev. D \textbf{52}, 4430 (1995).




%\bibitem{BTZ}M. Banados, C. Teitelboim and J. Zanelli, Phys. Rev. Lett. \textbf{72}, 957 (1994).


%\bibitem{GR}R. M. Wald, General Relativity, University of Chicago Press (Chicago, 1984).







%\bibitem{DSSY}
%N.~Deruelle, M.~Sasaki, Y.~Sendouda and D.~Yamauchi,
%``Hamiltonian formulation of f(Riemann) theories of gravity,''
%Prog. Theor. Phys. \textbf{123}, 169 (2010).


%\cite{Jiang:2018sqj}
%\bibitem{JZ}
%J.~Jiang and H.~Zhang,
%``Surface term, corner term, and action growth in $F(R_{abcd})$ gravity theory,''
%Phys. Rev. D \textbf{99}, 086005 (2019).
%48 citations counted in INSPIRE as of 14 Jun 2024
%\cite{Deruelle:2009zk}

%127 citations counted in INSPIRE as of 14 Jun 2024
%\cite{Harlow:2019yfa}

%\bibitem{BY}
%J. D. Brown and J. W. York, Phys. Rev. D \textbf{47}, 1407 (1993).

%\bibitem{MY}M. R. Visser and Z. Yan, arXiv: 2403.07140[hep-th].


%\bibitem{India}M. Maitra, D. Maity, and B. R. Majhi, Phys. Rev. D \textbf{97}, 124065 (2018).
%410 citations counted in INSPIRE as of 15 Jun 2024

%\bibitem{pa}T. Padmanabhan, Phys. Rev. D \textbf{84}, 124041 (2011).










%\bibitem{B}M. Banados and F Mendez, Phys. Rev. D 58, 104014(1998).














\end{thebibliography}
\end{document}